\documentstyle[aps,preprint,epsf]{revtex}

\begin{document}
\tightenlines
\draft

\title{Axial vector current in an electromagnetic field and \\
   low-energy neutrino-photon interactions} 

\author{H.~ Gies\thanks{
E-mail: holger.gies@uni-tuebingen.de
} and R.~ Shaisultanov\thanks{
E-mail: shaisul@pion14.tphys.physik.uni-tuebingen.de
}}

\address{Institut f\"{u}r theoretische Physik,
 Universit\"{a}t T\"{u}bingen,\\
 Auf der Morgenstelle 14, 72076 T\"{u}bingen, Germany }

\maketitle
\begin{abstract}
  An expression for the axial vector current in a strong, slowly
  varying electromagnetic field is obtained. We apply this expression
  to the construction of the effective action for low-energy
  neutrino-photon interactions.
\end{abstract}
\newpage
 \narrowtext
\section{Introduction}
The study of neutrino-photon interactions began many years ago when
Pontecorvo and Chiu and Morrison \cite{pcm} suggested that the process
$\gamma\gamma\to \nu\bar{\nu}$ may be important in the analysis of
stellar cooling.  These interactions are of interest in astrophysics
and cosmology.

It is well known that, in the standard model, neutrino-photon
interactions appear at the one-loop level. In addition to the heavy
gauge bosons which can immediately be integrated out at the desired
scales leading to the effective four-fermion interaction, charged
fermions (electrons) run in the loop; in particular, the coupling of
the photons to these fermions is responsible for the
$\gamma\nu$ interactions.

A variety of processes of physical relevance in different situations
belong to this class of $\gamma\nu$ interactions, for which an
effective-action description involving only neutrino currents and
field strength tensors has partly been found. E.g., Dicus and Repko
\cite{dr} have recently derived an effective action for interactions
between two neutrinos and three soft photons, which immediately allows
for the calculation of all scattering amplitudes of this type. This
effective action can be derived via the expectation value of the
electromagnetic vector current $\langle j_\mu\rangle = e\left\langle
\overline{\Psi}(x) \gamma _{\mu }\Psi (x)\right\rangle$; the latter
can in turn be calculated very easily from the Heisenberg-Euler
effective Lagrangian ${\cal L}_{\text{HE}}$ with the aid of the
formula $\left\langle \overline{\Psi}(x) \gamma _{\mu }\Psi
(x)\right\rangle = -{\delta {\cal L}_{\text{HE}}}/{e\: \delta A^{\mu
}(x)}$. In \cite{abada}, the effective action of Dicus and Repko was
reproduced by a more direct diagrammatic approach.

Incidentally, processes with two photons, for example, neutrino-photon
scattering \( \gamma \, \nu \rightarrow \gamma \, \nu \),
turn out to be highly suppressed in the standard model \cite{a,a2,b,c}
because, according to Yang's theorem \cite{yang}, two photons cannot
couple to the \( J=1 \) state. As a result, typical cross sections are
exceedingly small and suppressed by factors of \( 1/m_{W}^{2} \) (see,
e.g., \cite{f}). 

The presence of a medium or external electromagnetic field drastically
changes the situation. It induces an effective coupling between
photons and neutrinos, even enhancing processes such as $\nu
\rightarrow \nu \gamma$. Based on the same effective action of Dicus
and Repko \cite{dr}, it was shown in \cite{rashid} that in the
presence of an external magnetic field, cross sections for
neutrino-photon processes such as $\gamma\gamma\to \nu\bar{\nu}$ and
$\nu\gamma\to \nu\gamma$ are amplified by the factor \( \sim \left(
  m_{W}\, \, /m_{e}\right) ^{4}\left( B/B_{\text{c}}\right) ^{2} \)
for soft photon frequencies $\omega \ll m_{e}$ and a weak magnetic
field $B \ll B_{\text{c}}\equiv m^2/e$ (for extensions to this result,
see \cite{vas} and \cite{kit}). The subject of the electromagnetic
properties of neutrinos in media was comprehensively studied in
\cite{nieves} (see also \cite{grasso} for neutrinos in magnetized
media). 

In general, an effective action describing low-energy neutrino-photon
processes with an arbitrary number of photons and in the presence of
strong external electromagnetic fields appears highly desirable. In
other words, we are looking for the analogue of the Heisenberg-Euler
effective action which turned out to be extremely useful in QED (see
e.g., \cite{ditt}).

For this, we start from the effective four-fermion interaction
Lagrangian valid at energies very much smaller than the W- and Z-boson
masses:
\begin{equation}
\label{ha1}
{ \cal L } _{4f}=\frac{G_{F}}{\sqrt{2}}\overline{\nu }\gamma ^{\mu }
\left( 1+\gamma _{5}\right) \nu \overline{E}\gamma _{\mu }
\left( g_{V}+g_{A}\gamma _{5}\right) E.
\end{equation}
Here, $E$ denotes the electron field,
$\gamma_5=-i\gamma^0\gamma^1\gamma^2\gamma^3$,
$g_V=1-\frac{1}{2}\left(1-4\sin^2\theta_W\right)$ and
$g_A=1-\frac{1}{2}$ for $\nu_{e}$, where the first terms in $g_V$ and
$g_A$ are the contributions from the W exchange diagram and the second
one from the Z exchange diagram. Also,
$g_V=2\sin^2\theta_W-\frac{1}{2}$ and $g_A=-\frac{1}{2}$ for
$\nu_{\mu,\tau}$. 

Now concentrating on soft electromagnetic degrees of freedom with
energies much smaller than the electron mass, we may integrate out the
actual ``heaviest'' particle, i.e., the electron, and find
\begin{equation}
{ \cal L } _{\text{eff}}=\frac{G_{F}}{\sqrt{2}} \frac{1}{e}\,\,
\overline{\nu }\gamma ^{\mu } \left( 1+\gamma _{5}\right) \nu\,\, 
\bigl( g_{\text{V}} \langle j^\mu\rangle^A +g_{\text{A}} \langle
j_5^\mu\rangle^A \bigr) ,\label{Leff1}
\end{equation}
where the expectation values of the currents are given in terms of the
Green's function in this field: e.g., $\langle j_5^\mu\rangle^A= \text{i}
e\text{tr}\bigl[  \gamma^\mu \gamma_5\, G(x,x|A) \bigr]$.

Obviously, in order to obtain this effective Lagrangian, one must
calculate the expectation values of vector and axial vector currents
in a slowly varying electromagnetic field background. As mentioned
above, the vector current expectation value can be easily obtained
using the well-known Heisenberg-Euler Lagrangian ${\cal
  L}_{\text{HE}}$: $\left\langle \overline{E}(x)\gamma _{\mu }E
  (x)\right\rangle = -{\delta {\cal L}_{\text{HE}}}/{e\: \delta A^{\mu
    }(x)} $. In this way, one obtains a derivative expansion of the
vector current around a strong field.  For example, the term which is
third order in the field and first order in derivatives was used by
Dicus and Repko \cite{dr} in their study of $\nu \gamma \to \nu \gamma
\gamma$ and cross processes.

With regard to Eq. (\ref{Leff1}), the derivative expansion of the
axial vector current around an arbitrarily strong field is finally
required. To our surprise, we could not find such an expression in the 
vast literature on derivative expansions. Therefore, its derivation
remains the final task of our present work. 

It is convenient to choose a special gauge condition for the potential
$A_{\mu}(x)$ without loss of generality: the so-called
Fock--Schwinger gauge\footnote{Especially for gradient expansions of
  the heat kernel or the Heisenberg-Euler Lagrangian, this gauge has
  proved extremely useful; see, e.g., \cite{hauk84,gusy98}.}
\begin{equation}
(x^{\mu }-y^{\mu })A_{\mu }(x)=0,
\end{equation}
which is satisfied by the series
\begin{equation}
A_{\mu }(x)=\frac{1}{2}(x^{\lambda }-y^{\lambda })F_{\lambda \mu }(y)+
\frac{1}{3}(x^{\lambda }-y^{\lambda })(x^{\sigma }-y^{\sigma })
\partial _{\sigma }F_{\lambda \mu }(y)+ \ldots
\label{fock}
\end{equation}
Abbreviating the first term on the right-hand side with
$A_{\text{c}\mu}$ and the second term with $a_\mu$, we may perform a
perturbation expansion for the Green's function with respect to the
derivative term $a_\mu$:
\begin{equation}
G(x,x' |A)=G_{c}(x,x'|A_{\text{c}} )+
\int d^{4}w\, G_{c}(x,w|A_{\text{c}})\, ea_{\mu }(w)\,
\gamma ^{\mu}G_{c}(w,x'|A_{\text{c}} )+\ldots,
\label{green}
\end{equation}
where $G_{c}(x,x'|A_{\text{c}} )$ is the Green's function
of the electron in the constant field produced by $A_{\text{c}}$.
Inserting the expansion (\ref{green}) into the definition of $\langle
j^\mu_5\rangle^A$, we obtain, to first order in derivatives,
\begin{equation}
\langle j_5^\mu\rangle^A=\frac{1}{3}\partial_{\sigma}F_{\alpha
  \beta}\frac{\partial^2}{\partial k^{\sigma}\partial k^{\alpha}}
\Bigl[\Pi_5^{\beta \mu}(-k)\Bigr]\Big|_{k=0},
\label{ax}
\end{equation}
where we encounter the vector--axial-vector amplitude (VA)
$\Pi_5^{\beta\mu}$, i.e., the axial analogue of the polarization
tensor. This VA amplitude has been calculated very recently by one of
the authors in \cite{m5}; an independent confirmation can be found in
\cite{schubII}.

Inserting the presentation of \cite{m5} into Eq. (\ref{ax}), we
finally arrive at the first-order gradient expansion of the axial
vector current for arbitrarily strong electromagnetic fields valid for
slowly varying fields:
\begin{eqnarray}
\langle j_5^\mu\rangle^A&=& \text{i} e\, \text{tr}\, \bigl[\gamma^\mu 
\gamma_5\,  G(x,x|A)\bigr] \nonumber\\
&=& \text{i} \frac{e^{2}}{24\pi ^{2}}
\partial _{\sigma }F_{\lambda \nu }
\int ^{\infty }_{0}\frac{ds}{s}\, e^{-ism^{2}}\frac{(eas)(ebs)}
{\sin (ebs)\sinh (eas)}\label{current}\\
 &  & \left\{ \left[ C^{\nu \lambda }\left( C^{2}\right) ^{\mu \sigma
 } +C^{\nu \sigma }\left( C^{2}\right) ^{\mu \lambda }\right] N_{1}+
\left[ B^{\nu \lambda }\left( B^{2}\right)^{\mu\sigma}+B^{\nu\sigma}
\left( B^{2}\right) ^{\mu \lambda }\right] N_{2}\right. \nonumber\\
 &  &  \qquad -\left.\left[ 3F^{*\nu \mu }g^{\lambda \sigma }+
F^{*\nu \lambda }g^{\mu \sigma }+F^{*\nu \sigma }
 g^{\mu \lambda }\right] N_{3} \right\}, \nonumber
\end{eqnarray}
where the $N_i$ are simple functions of the field strength:
\begin{eqnarray}
N_{1}=\frac{2\sin (ebs)}{\sinh ^{2}(eas)}\left[ \cosh (eas)-
\frac{\sinh (eas)}{eas}\right] +2ebs\,N_{3} &  &\nonumber \\
N_{2}=\frac{2\sinh (eas)}{\sin ^{2}(ebs)}\left[ -\cos (ebs)+
\frac{\sin (ebs)}{eb
s}\right] +2eas\,N_{3} &  &\nonumber \\
N_{3}=\frac{1}{eas\,ebs(a^{2}+b^{2})}\left[ a^{2}\frac{\sin (ebs)}
{\sinh (eas)}-b^{2}\frac{\sinh (eas)}{\sin (ebs)}\right].  &  & \nonumber
\end{eqnarray}
For the analysis, we employed a decomposition of the field strength
tensor in linearly independent subspaces of electric and magnetic parts:
\begin{equation}
\label{ha65}
C_{\mu \nu }:=\frac{\displaystyle 1}{\displaystyle a^{2}+b^{2}}
\left( aF_{\mu \nu }+bF^{*}_{\mu \nu }\right)\, ; \quad  
B_{\mu \nu }:=\frac{\displaystyle 1}{\displaystyle a^{2}+b^{2}}
\left( bF_{\mu \nu }-aF^{*}_{\mu \nu }\right)\, , 
\end{equation}
where the invariants $a$ and $b$ are defined by
\begin{equation}
\label{ha4}
a,b=\sqrt{({\cal F}^{2}+{\cal G}^{2})^{\frac{1}{2}}\pm 
{\cal F}};\quad  {\cal F}=-\frac{1}{4}F_{\mu \nu }F^{\mu \nu },\quad
{\cal G}=-\frac{1}{4}F^{*}_{\mu \nu }F^{\mu \nu } .
\end{equation}
As a consequence, we find\footnote{We always use
the metric $g={\rm diag}({+}{-}{-}{-})$.}
\begin{equation}
\label{ha655}
\left( C^{2}\right) _{\mu \nu }=\frac{\displaystyle 1}
{\displaystyle a^{2}+b^{2}}\left( F^{2}_{\mu \nu }+b^{2}
g_{\mu \nu }\right) ;\quad  \left( B^{2}\right) _{\mu \nu }=
\frac{\displaystyle 1}{\displaystyle a^{2}+b^{2}}
\left( F^{2}_{\mu \nu }-a^{2}g_{\mu \nu }\right)\, .
\end{equation}
Concerning our central result for the axial vector current in
Eq. (\ref{current}), we observe its linearity in the derivative of
$F_{\mu\nu}$ (by construction) which is multiplied by a finite
proper-time integral containing the complete dependence on the
strength of the fields. We would like to point out that our method of
performing the derivative expansion in momentum space
(cf. Eq. (\ref{ax})) is much in the spirit of Ref. \cite{vain} with
the essential difference that we expand around an arbitrary value of
field strength. 

Our representation of $\langle j_5^\mu\rangle$ in terms of the
$C_{\mu\nu}$ and $B_{\mu\nu}$ tensors is very convenient for
performing a weak-field expansion; to second order in $1/m^2$, it
finally reads: 
\begin{eqnarray}
\langle j_{5}^{\mu }\rangle ^{A}&=&\frac{\displaystyle
e^{3}}{\displaystyle 24\pi ^{2}m^{2}}
\Bigl (\partial^{\mu} {\cal G}+(\partial ^{\alpha }F_{\alpha \beta })
F^{*\beta \mu }\Bigr ) \label{weakf}\\
&&+\frac{\displaystyle e^{5}}{\displaystyle 90\pi ^{2}m^{6}}
\partial _{\sigma }F_{\alpha \beta }
\Bigl [{\cal G} \bigl(F^{\beta \alpha }g^{\mu \sigma }+
F^{\beta\sigma} g^{\mu\alpha}\bigr)
+ \bigr( F^{*\beta \alpha }(F^{2})^{\mu \sigma }
     + F^{*\beta\sigma} (F^2)^{\mu\alpha}\bigr) \nonumber\\
&&\qquad\qquad\qquad\qquad - {\cal F}\bigl(3F^{*\mu \beta }g^{\alpha
  \sigma }+  F^{*\beta \alpha }g^{\mu \sigma }+
F^{*\beta \sigma }g^{\mu \alpha }\bigr)\Bigr ]. \nonumber 
\end{eqnarray}
As a cross-check, one can indeed prove \cite{HGRS1} that the first
term can immediately be obtained from the order $1/m^2$ term of the
famous triangle graph. As a side remark, it should be mentioned that
the axial vector anomaly is not present in our result, since we
computed the $1/m^2$-expansion of $\langle j_5^\mu\rangle$, whereas
the anomaly is mass independent.

Returning to the quest for the general effective action, we can easily
insert Eq. (\ref{current}) into Eq. (\ref{Leff1}) and a similar
expression for the vector current $\langle j_\mu\rangle$. The
well-known latter expression can be derived via the Heisenberg-Euler
Lagrangian as mentioned above or, alternatively, via the same method
as here proposed for the axial vector current; for this, we would like
to stress that Eq. (\ref{ax}) also holds for the vector current with
the VA amplitude $\Pi_5^{\mu\nu}$ replaced by the polarization tensor
$\Pi^{\mu\nu}$. This tensor in the presence of an arbitrarily constant
electromagnetic field was first calculated in \cite{shabad} (an
appropriate representation can be found in \cite{urru79}). Details
will be presented elsewhere. We observe that the axial vector current
is of even order in the field, while the vector current is of odd
order; this is a direct consequence of the Dirac algebra (Furry's
theorem).

To conclude, we have completed the search for a low-energy effective
action for neutrino interactions with an arbitrary number of soft
photons with wavelengths larger than the Compton wavelength. From a
different perspective, we have found the generalization of the
Heisenberg-Euler action to the case involving an axial vector
coupling.

\section*{Acknowledgments}

We would like to thank Professor W. Dittrich for helpful discussions
and for carefully reading the manuscript. This work was supported by
Deutsche Forschungsgemeinschaft under grant DFG Di 200/5-1.

\end{document}